\documentclass[fleqn,twoside]{article}
\usepackage{amsmath}
\usepackage{espcrc2}
\usepackage{graphicx}

\newcommand{\Msun}{\mbox{ M$_\odot$}}
\newcommand{\Mpc}{\mbox{ Mpc}}
\newcommand{\Mhz}{\mbox{ MHz}}

\newcommand{\kel}{\mbox{ K}}
\newcommand{\km}{\mbox{ km}}
\newcommand{\mkel}{\mbox{ mK}}
\newcommand{\fluxden}{\mbox{ erg cm$^{-2}$  s$^{-1}$ Hz$^{-1}$ sr$^{-1}$}} 
\newcommand{\beq}{\begin{equation}}
\newcommand{\eeq}{\end{equation}}
\newcommand{\beqa}{\begin{eqnarray}}
\newcommand{\eeqa}{\end{eqnarray}}

\newcommand{\xh}{{x_H}}
\newcommand{\bxh}{{{\bar x}_H}}

\title{21 cm Tomography of the High-Redshift Universe with the Square
  Kilometer Array}

\author{S. R. Furlanetto\address{Division of Physics, Mathematics, \&
        Astronomy, California Institute of Technology, \\ Mail Code
        130-33, Pasadena, CA 91125, USA}
        and F. H. Briggs\address{Research School of Astronomy \& Astrophysics,
        Mount Stromlo Observatory, Cotter Road, Weston, ACT 2611, Australia}
}
       
\begin{document}

\begin{abstract}

We discuss the prospects for ``tomography'' of the intergalactic
medium (IGM) at high redshifts using the 21 cm transition of neutral
hydrogen.  Existing observational constraints on the epoch of
reionization imply a complex ionization history that may require
multiple generations of sources.  The 21 cm transition provides a
unique tool to probe this era in detail, because it does not
suffer from saturation effects, retains full redshift information, and
directly probes the IGM gas.  Observations in the redshifted 21cm line will
allow one to study the history and morphology of reionization in
detail.  Depending on the characteristics of the first sources, they
may also allow us to probe the era before reionization, when the first
structures and luminous sources were forming.  The construction of
high signal-to-noise ratio maps on arcminute scales will require
approximately one square kilometer of collecting area.

\end{abstract}

\maketitle

\section{Introduction}

One of the major goals of modern cosmology is to understand how the
first luminous sources in the universe formed and grew into the
galaxies and galaxy clusters that we see today.  As technology has
improved in the past four decades, the frontiers of our knowledge have
extended farther and farther into the past.  We now sit at the cusp of
exploring the era in which the first generations of sources were just
beginning to form:  the end of the cosmological ``dark ages.''
Observations of this epoch will teach us about the transformation from the
smooth, relatively simple universe that we see in the cosmic microwave
background (CMB) to the complex universe that we see nearby.

The hallmark of the era of first objects is the ``reionization'' of
the intergalactic medium (IGM), when these first luminous sources
ionized the gas around them and rendered it transparent to
ultra-violet photons that we observe today as optical or infrared
light (see \cite{barkana01} and references therein).  Fortunately,
radio photons, including those from the CMB, readily propagate through
both the neutral and ionized IGM at this time, and these will help us
to measure an enormous amount of information about the process of
reionization, the properties of the first sources, and the properties
of the IGM \cite{wyithe03,cen03,haiman03}.

Existing techniques have provided tantalizing constraints on the epoch
of reionization (EOR).  The most straightforward method is to observe
the high redshift part of the ``Ly$\alpha$ forest:'' regions with
relatively large HI densities appear as absorption troughs in quasar
spectra, which presumably deepen and come to dominate the spectra as
we approach the reionization epoch.  Indeed, spectra of $z \sim 6$
quasars selected from the Sloan Digital Sky Survey (SDSS) show at
least one extended region of zero transmission \cite{becker},
indicating that the ionizing background is rising rapidly at this time
\cite{fan} (but see \cite{songaila04}).  However, the mean optical
depth of the IGM to Ly$\alpha$ absorption is $\tau_{Ly\alpha} \approx
6.45 \times 10^5 \xh [(1+z)/10]^{3/2}$ \cite{gp}, where $\bxh$ is the
global neutral fraction.  Even a modest $\bxh > 10^{-3}$ will
therefore render the absorption trough completely black; quasar
absorption spectra can clearly probe only the latest stages of
reionization.

A second constraint comes from the effects of the ionized gas on the
CMB.  The free electrons Thomson scatter CMB photons, washing out
the intrinsic anisotropies but generating a faint polarization signal.
The total scattering optical depth $\tau_{\rm es}$ is proportional to
the column density of ionized hydrogen, so it provides an integral
constraint on the reionization history.  Recently, the \emph{Wilkinson
Microwave Anisotropy Probe} (\emph{WMAP}) used the polarization signal
to measure a large $\tau_{\rm es}$, indicating that reionization began
at $z > 14$ \cite{kogut03,Spergel03}.  A third (possibly conficting)
constraint comes from measurements of the temperature of the
Ly$\alpha$ forest at $z \sim 2$--$4$, which, taken at face value,
require reheating of the IGM at $z<10$.  Since heating is accomplished
through photoionization, this would suggest an order unity change in
the ionized fraction at $z < 10$ \cite{theuns02-reion,hui03}, though
with large uncertainties.

\begin{figure}[htb]
\includegraphics[width=3in]{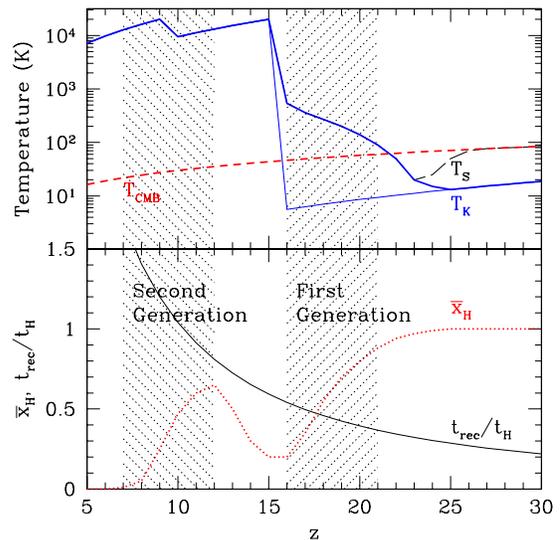}
\vspace{-0.5in}
\caption{A cartoon illustrating a double reionization history.  The
  bottom panel shows the ratio of the recombination time and the age
  of the universe (solid curve) as well as $\bxh$ (dotted curve).  The
  top panel shows the kinetic temperature of the IGM gas (solid
  lines), the CMB temperature (short-dashed line), and the spin
  temperature (long-dashed line, note that for $z<23$ we assume this
  traces $T_K$).  The thick solid line neglect all heating mechanisms
  except for photoionization, while the thin line shows what could
  happen if X-ray and shock heating were included (see \S 3).  The
  shaded regions mark the two generations of ionizing sources.}
\label{fig:cartoon}
\end{figure}

Taken together, these three sets of observations imply a complex
reionization history extending over a large redshift interval ($\Delta
z \sim 10$).  This is inconsistent with a ``generic'' picture of fast
reionization (e.g., \cite{barkana01}, and references therein).  The
results seem to indicate strong evolution in the sources responsible
for reionization, and a detailed measurement of the reionization
history would contain a rich set of information about early structure
formation \cite{sokasian03a,wyithe03,cen03,haiman03}.  In fact, it is
quite plausible that the ionization history is complex, with long
periods of ``stalling'' or even two distinct epochs of reionization.
Figure~\ref{fig:cartoon} illustrates why.  Here we consider the
possibility of two generations of ionizing sources
\cite{wyithe03,cen03}.  In the bottom panel we show the ratio of the
recombination time (assuming the mean cosmic density and $T=10^4
\kel$, typical of ionized gas) to the age of the universe (solid
curve).  Note that the ratio is near unity at $z \sim 10$ and
decreases at earlier times, implying that a significant fraction of
the gas can recombine if it is ionized by this point.  The dotted
curve shows how $\bxh$ could evolve and roughly satisfy the three
observational constraints.  The first generation of sources (for
example, very massive metal-free stars) ionize most of the gas.
However, these sources quickly cease forming (perhaps because they
enrich their environs with heavy elements) and most of the gas
recombines.  Full reionization must await the second generation of
sources (in our example, Population II stars).  We see from the Figure
that the 21 cm structures could be long-lived and complex: for
example, gas above the mean density may recombine between the two
generations while voids remain ionized.  The top panel illustrates the
temperature evolution of the IGM in this scenario; note that
reionization corresponds to a sudden boost in the temperature.  This
panel will be discussed in more detail below.

While such cartoon views are certainly plausible, detailed models rely
on a number of uncertain parameters that we would like to measure.
The optimal reionization experiment would: (1) be sensitive to order
unity changes in $\bxh$ (to probe the crucial middle stages of
reionization), (2) provide measurements that are well-localized along
the line of sight (rather than a single integral constraint), and (3)
not require the presence of bright background sources, which may be
rare at high redshifts.  The most promising candidate proposed to date
is the 21 cm hyperfine transition of neutral hydrogen in the IGM
\cite{field58,field59a}, which fulfills all three of these criteria.

So long as the excitation temperature $T_S$ of the 21 cm transition in
a region of the IGM differs from the CMB temperature, that region will
appear in either emission (if $T_S > T_{\rm CMB}$) or absorption (if
$T_S < T_{\rm CMB}$) when viewed against the CMB.  Variations in the
density of neutral hydrogen (due either to large-scale structure or to
HII regions) would appear as fluctuations in the sky brightness of
this transition.  Because it is a line transition, the fluctuations
can also be well-localized in redshift space.  Thus, in principle,
high resolution observations of the 21 cm transition in both frequency
and angle can provide a three-dimensional map of reionization.
Together with radio absorption spectra of bright background sources
(see the contribution by Carilli, this volume), these observations promise
to shed light both on the early growth of structure and on
reionization.

In the following, we will distinguish two observational goals.  The
ultimate goal is to make three-dimensional, high signal-to-noise
\emph{maps} of the sky in the 21 cm transition.  In addition to
precisely mapping reionization itself, this would, for
example, allow detailed comparisons between the distribution of
neutral gas and the ionizing sources.  Unfortunately, we will find
below that the expected brightness of the IGM is extremely small both
in absolute terms and when compared to the relevant foreground
contaminants.  It is therefore considerably easier to make
\emph{statistical} measurements of the distribution of neutral gas.
As we have seen with CMB measurements, these statistical constraints
can be quite powerful.

In \S 2 we will briefly review the important physics of the 21 cm
transition.  In \S 3 we will outline the information we can expect to
learn from 21 cm observations, and in \S 4 we will describe the
instrumental requirements for making maps and statistical
measurements.  Finally, in \S 5 we conclude and discuss 21 cm
observations in relation to other instruments that will be built in
the next decade.

\section{The 21 cm Transition}

We define $\delta T(\nu)$ to be the observed brightness temperature
increment between a patch of the IGM, at a frequency $\nu$
corresponding to a redshift $1+z=\nu_0/\nu$ (where $\nu_0=1420.4 \Mhz$
is the rest frequency of the 21 cm transition), and the CMB.  Assuming the
\emph{WMAP} cosmological parameters \cite{Spergel03} and small 21 cm
optical depth $\tau_{21}<<1$, this quantity is
\cite{Scott90,Mad97}: 
\beqa 
\delta T(\nu) & \approx & \frac{T_S - T_{\rm CMB}}{1+z} \, \tau_{21}
\label{eq:dtb} \\
\, & \approx \, & 23 \, (1+\delta) \xh {\mathcal T} 
\left( \frac{1+z}{10} \right)^{1/2} \mkel,
\nonumber
\eeqa
where ${\mathcal T} = (T_S - T_{\rm CMB})/T_S$, $T_S$ is the spin
temperature of the IGM, $\xh$ is the neutral fraction, and $\delta$ is
the overdensity.  All of these local quantities should be averaged
over the volume sampled by the telescope beam (note that we will use
$\bxh$ for the global, mass-averaged neutral fraction).  For
reference, one arcminute subtends $\approx 1.9 h^{-1}$ comoving Mpc at
$z=10$ in the WMAP cosmology, while a bandwidth of $0.1 \Mhz$
corresponds to $\approx 1.7 \Mpc$ at the same redshift.

The observability of the IGM clearly depends on the spin temperature,
which may be written \cite{field58,field59a}:
\beq
T_S = \frac{T_{\rm CMB} + y_c T_K + y_{{\rm Ly}\alpha} T_{{\rm
      Ly}\alpha}}{1 + y_c +  y_{{\rm Ly}\alpha}}. 
\label{eq:hItspin}
\eeq 
The second term describes collisional excitation of the hyperfine
transition, which couples $T_S$ to the gas kinetic temperature $T_K$.
The coupling coefficient $y_c \propto n_H$ has been numerically
evaluated by \cite{allison}; it becomes strong when $1+\delta > 5
[(1+z)/20]^2$ and is not important in the smooth IGM in the relevant
redshift range.  The third term in equation (\ref{eq:hItspin})
describes the Wouthuysen-Field effect, in which Ly$\alpha$ pumping
couples the spin temperature to the color temperature of the radiation
field $T_{{\rm Ly}\alpha}$ \cite{wout,field58} (which equals $T_K$ in
most astrophysical environments \cite{field59b}).  The physics of this
mechanism have been described recently in \cite{Meik99}.  Ly$\alpha$
pumping effectively couples $T_S$ and $T_K$ when the radiation
background at the Ly$\alpha$ line center satisfies $J_\alpha >
10^{-21} \fluxden$.  In a low-density, neutral IGM, such a background
would need to be present in order for the IGM to be visible in the 21
cm transition.

\section{A Short History of the Universe}

It is obvious from the above discussion that the IGM brightness
temperature $\delta T(\nu)$ depends sensitively on the thermal history
of the IGM.  The spin temperature and optical depth therefore depend
on the kinetic temperature and density of the IGM, as well as the mean
intensity of the Ly$\alpha$ flux.  Once Thomson scattering of CMB
photons becomes inefficient at the thermal decoupling redshift $z_d
\sim 140$, the IGM cools adiabatically until the first objects
collapse.  During this era, $T_K<T_{\rm CMB}$; observations at $\nu <
30 \Mhz$ could in principle detect the IGM in absorption during this epoch
\cite{loeb-21}.  The cooling trend reverses itself as structure begins
to form, but the subsequent temperature evolution is both
inhomogeneous and highly uncertain.  While early estimates suggested
that Ly$\alpha$ photons themselves would inject significant thermal
energy into the IGM, \cite{chen03} showed that this heating channel is
in reality quite slow.  Instead, X-rays (primarily from supernovae or
accreting black holes), adiabatic compression, and structure formation
shocks are likely to control the temperature evolution of the IGM.
Because of this complex spin temperature evolution, 21 cm signatures
can be divided into three separate eras.  Here we will describe each
of these phases in turn.  The top panel of Figure~\ref{fig:cartoon}
shows a cartoon of the qualitative temperature evolution in a double
reionization model.  To fix ideas, we will refer to this Figure~in the
following discussion.  We show two examples for $T_K$.  The thin line
includes only adiabatic cooling and photoionization heating (which we
have applied at the maximal ionized fraction from each generation for
simplicity).  The thick line has additional heat sources before
overlap.

\begin{figure*}[htb]
\includegraphics*[width=6in]{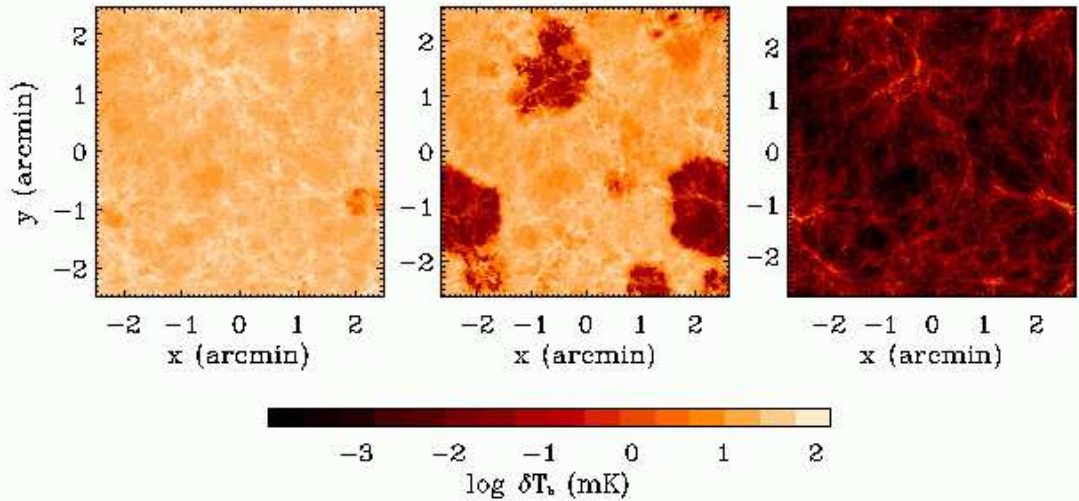}
\vspace{-0.5in}
\caption{ The brightness temperature of the 21 cm transition at
several redshifts, as predicted by the ``late reionization''
simulation analyzed in \cite{furl-21cmsim}.  Each panel corresponds to
the same slice of the simulation box (with width $10 h^{-1}$ comoving
Mpc and depth $\Delta \nu=0.1 \Mhz$), at $z=12.1$, 9.2, and 7.6, from
left to right.  The three epochs shown correspond to the early,
middle, and late stages of reionization in this simulation; we assume
$T_S >> T_{\rm CMB}$ throughout.}
\label{fig:sim}
\end{figure*}

\subsection{The First Structures}

The first phase occurs while luminous sources are rare; radiative
feedback (including both heating and Ly$\alpha$ coupling) can
therefore be ignored in most of the IGM.  Although $T_K<T_{\rm CMB}$
in this phase, we thus have $T_S=T_{\rm CMB}$ and most of the gas
remains invisible.  In Figure~\ref{fig:cartoon}, this era corresponds
to $z > 25$.  However, note that only moderate overdensities are
required in order to force $T_S$ and $T_{\rm CMB}$ to depart from each
other.  One regime in which collisions are important are
``minihalos'': collapsed objects that are not massive enough to cool
and form stars.  With mean overdensities $\delta \sim 200$, these
objects are dense enough for collisions to quickly force the spin
temperature to the virial temperature of the gas.  They would
therefore appear in emission against the CMB.  While an individual
minihalo (with $M \sim 10^6 \Msun$) is much too small to be observed
with the SKA, the integrated signal from the distribution of minihalos
will cause \emph{rms} fluctuations of order $\delta T \sim 0.1$--$1
\mkel$ on angular scales of a few arcminutes \cite{Iliev02}, depending
sensitively on $z$.

However, as described in \S 2, collisional coupling does not require
virial densities.  More moderate overdensities, such as those that
appear during the collapse of proto-sheets and filaments, suffice to
heat the gas and force $T_S \rightarrow T_K$.  While these shocks are
difficult to model without detailed numerical simulations,
semi-analytic estimates suggest that shocked gas can amplify the
signal expected from minihalos by a factor of a few at $z\sim20$
\cite{furl-shock}.  Finally, bubbles of emission or absorption may
appear around the first luminous sources because their radiation
fields can trigger the Wouthuysen-Field mechanism locally
\cite{Mad97}.  Thus, in this phase we expect relatively weak
fluctuations to trace the formation of the first baryonic structures
in the universe.

\subsection{First Light}

This first phase ends when luminous sources become sufficiently
numerous so as to make the Wouthuysen-Field effect important globally.
At this point $T_S \rightarrow T_K$ in the diffuse IGM and it becomes
visible against the CMB \cite{Mad97}.  Assuming that the Ly$\alpha$
background turns on faster than the X-ray background, most of the IGM
will still be cold and will initially be visible in absorption.  In
the cartoon of Figure~\ref{fig:cartoon}, this era begins at $z \sim
25$ as the spin temperature falls below $T_{\rm CMB}$.  Superposed on
the absorbing background, shock-heated filaments and minihalos will
still appear in emission.  Thus in this epoch the 21 cm fluctuations
depend on the density, temperature, and ionized fraction.  The
situation is sufficiently complex that predictions require detailed
numerical simulations (to trace the temperature structure from shocks
in the IGM) as well as detailed modeling of the radiation background
(to describe the Ly$\alpha$ coupling and X-ray heating).  Such
predictions are not yet available on the large (several Mpc) scales
relevant for 21 cm observations, although simulations on
smaller scales are now becoming available \cite{Gnedin03}.
Nevertheless, this phase is particularly interesting because it
signals the appearance of the first luminous sources.  For example, as
shock-heating continues and the X-ray background builds, the diffuse
IGM will be heated above $T_{\rm CMB}$ and the mean absorption will
shift to emission, as occurs at $z \sim 21$ in
Figure~\ref{fig:cartoon}.  Comparing the onset of Ly$\alpha$ coupling
and the onset of heating will thus constrain the characteristics of
the first luminous sources.

\subsection{A Uniformly Heated IGM}

Once the Ly$\alpha$ and X-ray backgrounds become sufficiently strong,
the 21 cm fluctuation pattern becomes independent of the temperature
field.  Even though the IGM will likely have a complex temperature
field from continued shock heating, the brightness temperature is
essentially unaffected: ${\mathcal T} \rightarrow 1$ in equation
(\ref{eq:dtb}).  The 21 cm signal therefore depends only on $\delta$
and $\xh$ during this phase.  As a result, this epoch is more
straightforward to consider from a theoretical standpoint than those
described above.  Several groups have analyzed it recently with both
numerical \cite{Ciardi03,furl-21cmsim,Gnedin03} and analytic
\cite{Zald04,furl-21cmps,furl-21cmobs} techniques.  In Figure
\ref{fig:cartoon}, this approximation is accurate for all $z < 20$ or
so.  An example of the evolving fluctuation pattern during this phase
is shown in Figure~\ref{fig:sim}.

The ultimate observational goal of the SKA is to map the evolving
structure seen in the central panel of Figure~\ref{fig:sim}.
Structures in the IGM are causally related to the collapsed objects
that are responsible for the photoionizing flux, and measuring these
interelations will teach us about the first generations of stars and
protogalaxies.  However, direct imaging of the IGM in the 21cm line is
a formidable challenge, and it is likely that the first steps in
observing the 21cm signal from the EOR will be statistical studies
that discriminate between evolutionary scenarios through measurements
of (for example) the angular power spectrum.

\begin{figure}[htb]
\includegraphics[width=3in]{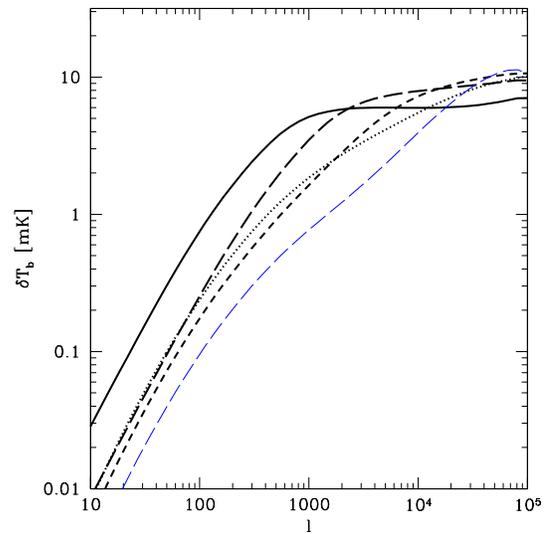}
\vspace{-0.5in}
\caption{An example of the time evolution of the 21 cm power spectrum.
  The time ordering is: dotted ($z=18$, $\bxh=0.96$), short-dashed
  ($z=15$, $\bxh=0.81$), long-dashed ($z=13$, $\bxh=0.52$) and solid
  ($z=12$, $\bxh=0.26$).  The thick curves assume the reionization
  model of \cite{furl-21cmps} while the thin dashed curve has each
  galaxy host an independent HII region ($z=13$, $\bxh=0.52$). The
  model assumes that each baryon inside a star forming galaxy produces
  $\zeta=40$ ionizing photons. From \cite{furl-21cmps}.}
\label{fig:ps}
\end{figure}

If uniform heating occurs sufficiently early compared to reionization,
fluctuations in $\xh$ will also be negligible.  In this case the 21 cm
field traces the density and yields a direct measurement of the power
spectrum of density fluctuations at $z \sim 10$ \cite{Tozzi00}.  Such
an observation would provide a sensitive probe of the growth of
fluctuations during the early universe.  The scales accessible to
observations are linear at this time, so predictions of the
fluctuation spectrum are relatively easy to make
\cite{Tozzi00,Zald04}.  Zaldarriaga et al. pointed out the
mathematical analogy between measurements of CMB fluctuations and the
21 cm pattern, with the key difference that (because it is a line
transition), 21 cm maps contain independent information at different
frequencies.  They showed how to describe the 21 cm fluctuations in
terms of their angular power spectrum, which can in turn be computed
from the three-dimensional power spectrum $P({\bf k})$.  The
fluctuations on a scale $l=2\pi/\theta$ are parameterized by the $C_l$
coefficients; the \emph{rms} fluctuations are then given by $\langle
\delta T^2 \rangle^{1/2} = l(l+1)C_l/2\pi$ (note 1 arcmin corresponds
to $l \approx 2 \times 10^4$).  In the case in which density
fluctuations dominate, this is simply the usual linear power spectrum
on the relevant scales.  The fluctuations increase with decreasing
angular scale, with a characteristic \emph{rms} magnitude of several
mK.  The dotted curve in Figure~\ref{fig:ps} shows the angular power
spectrum for this case of a universe with early, uniform heating at
$z=18$ and with $\bxh =0.96$. This spectrum essentially registers the
density fluctuations at $z=18$ and has no strong features.

Once luminous sources become important, the primary source of
fluctuations are the HII regions around ionizing sources, as shown in
the middle panel of Figure~\ref{fig:sim}.  Predictions of the
fluctuation pattern therefore depend on the particular model of
reionization.  One approach is to use numerical simulations to follow
reionization.  This is difficult for several reasons.  First,
radiative transfer remains a numerically challenging problem.  It has
usually only been added as a ``post-processing'' step to existing
hydrodynamic simulations.  Second, each simulation represents a
substantial investment of computer time, and it is presently
impractical to examine a large parameter space of reionization models.
Third, locating typical ionizing sources (with $M \sim 10^8 \Msun$) at
these redshifts requires high mass resolution, limiting the size of
the computational box to $L < 10 h^{-1}$ comoving Mpc
\cite{sokasian03a}.  This severely limits the ability of numerical
simulations to predict the 21 cm pattern on the arcminute scales
(corresponding to several comoving Mpc) that would be accessible to
the SKA.  Nevertheless, much can be learned about the fluctuation
pattern by examining simulations on small scales and extrapolating to
larger scales \cite{Ciardi03,furl-21cmsim,Gnedin03}.  These studies
again suggest \emph{rms} fluctuations $\delta T \sim 1$--$10 \mkel$ on scales
of $\sim 1$--$10$ arcminutes ($l\sim 10^3$--$10^4$).  The fluctuations
also depend on the spectral bandwidth of the observation; they begin
to be significantly suppressed when the radial scale (set by the
spectral channel resolution) exceeds the transverse scale
\cite{Zald04,furl-21cmps}.  Thus the relevant angular scales
correspond to $\Delta \nu \sim 0.2$--$2 \Mhz$, or $\Delta z \sim
0.02$--$0.2$.

Analytic models present a different set of problems.  Most
importantly, reionization is a highly nonlinear process, and
constructing an accurate model for the sizes and spatial distribution
of HII regions is challenging.  The simulations described above all
show that the usual approximation -- single HII regions surrounding
each galaxy -- does not describe overlap well.  Instead, dense regions
of the universe containing many galaxies rapidly grow into large
ionized regions (with scales of a few comoving Mpc) once the ionizing
flux escapes from the over-densities into the low density regions.
Thus analytic treatments must take into account large scale features
of the density field and not simply the regions near single galaxies.
Moreover, there are a number of uncertain parameters (such as the
clumpiness of gas in the IGM) that must be calibrated to simulations.
Nevertheless, they offer the advantage of allowing efficient parameter
studies and are not limited to small scales.  Recently,
\cite{furl-21cmps} have developed an analytic model of reionization
that includes the crucial features of the reionization process.  The
model relies on a simple extension of the Press-Schechter model
\cite{Press74} to compute the ionization pattern.  It assumes that
each galaxy can ionize a fixed multiple $\zeta$ of its own baryonic
mass.  It then determines the sizes of HII regions by finding those
regions that contain enough collapsed gas (i.e., galaxies) to ionize
the entire region; \cite{furl-21cmps} showed that this can be
approximated relatively simply through the ``excursion set'' formalism
\cite{bond91}.  Given the background cosmology, the size distribution
depends on only two parameters: the number of ionizations per
collapsed baryon ($\zeta$), and the minimum halo mass that can host an
ionizing source.

\begin{figure}[htb]
\includegraphics[width=3in]{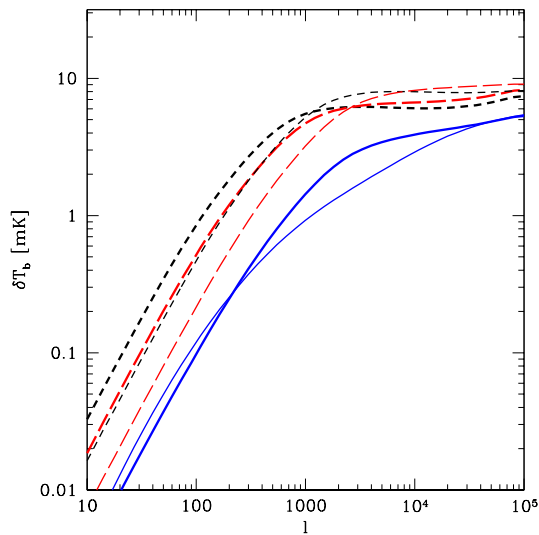}
\vspace{-0.5in}
\caption{ The 21 cm power spectrum in three different models of
  ``double reionization.''  The solid curves assume that the universe
  was uniformly ionized to $\bxh=0.5$ by a first generation of
  objects.  The others assume that the first generation shut off while
  discrete bubbles still existed, again when $\bxh=0.5$.  The second
  generation of sources completes reionization in all cases.  In each
  model, the thin curves have a total $\bxh=0.46$ and the thick curves
  have $\bxh=0.27$.  From \cite{furl-21cmobs}.}
\label{fig:ps2}
\end{figure}

Given a model for reionization, we can then determine the observable
characteristics of a 21 cm measurement.  With high signal-to-noise
maps, the size distribution of HII regions and the characteristics of
reionization can be measured in a straightforward way.  However, even
without maps, statistical measurements are still extremely powerful.
As described above, the simplest approach is to compute the power
spectrum of 21 cm emission, which depends on both the size
distribution of HII regions and fluctuations in the background density
field.  It is also relatively easy to interpret.  Several examples of
this simple statistic are shown in Figures \ref{fig:ps} and
\ref{fig:ps2} (both are taken from \cite{furl-21cmps}).
Figure~\ref{fig:ps} shows how the power spectrum evolves with time as
the global neutral fraction decreases from unity to $\bxh =0.26$.  As
the HII regions appear and grow, they add a feature to the power
spectrum on scales of several arcminutes ($l \sim 10^3$--$10^4$).  The
\emph{rms} fluctuations increase by a factor of several on the characteristic
scale of the HII regions, yielding a clear signature of reionization
even in statistical measurements.  The crucial point is that the power
spectrum evolves rapidly through reionization, so its measurement will
allow us to trace the time history of the HII regions.  The thin curve
shows an example where the sizes of HII regions are determined by
individual galaxies (as if the galaxies were randomly distributed); it
emphasizes the importance of the model of reionization in predicting
the power spectrum features and, conversely, the power of 21 cm
tomography to differentiate these models.  Note that these Figures
assume infinite frequency resolution; as discussed above, the
fluctuations are significantly suppressed if the comoving radial
distance exceeds the transverse distance \cite{Zald04,furl-21cmps}.
Until late in reionization (when the HII regions become very large),
the desired bandwidths correspond to $\Delta z \sim 0.02$--$0.2$, and
the redshift evolution within each bin is modest.

Figure~\ref{fig:ps2} shows the kind of detailed information that we
can extract from 21 cm tomography through the power spectrum
\cite{furl-21cmobs}.  Here we contrast predictions for two different
prescriptions of ``double reionization,'' a scenario first considered
in \cite{cen03,wyithe03}.  Specifically, we consider the mechanism
through which the first generation of highly ionizing sources shuts
off.  One possibility (solid curves) is that these sources ionize the
universe more or less uniformly (or that they fully ionize the
universe, which then recombines to a nearly uniform neutral fraction).
Another possibility is that the first generation shuts off while its
HII regions still exist, so that the neutral fraction is highly
inhomogeneous during the transition.  The two sets of dashed curves in
Figure~\ref{fig:ps2} show this prescription, with two different sets
of properties for the first generation of sources.  We see that, in
the former case, the power spectrum and the bubble feature are
significantly suppressed.  In the latter scenario, however, the bubble
feature is quite strong.  Essentially, the first generation imprints a
characteristic bubble size that persists until the second generation
of sources are able to dominate the total number of ionizations.
While these scenarios are no doubt overly simplified, they serve to
illustrate the power of 21 cm observations.

It is important to note, however, that the power spectrum is only the
simplest statistic available, essentially measuring the \emph{rms}
deviation as a function of angular scale.  There are several other
ways to approach the problem that take expicit advantage of the
three-dimensional nature of the observations \cite{Mor04}.  A related
question is how to take advantage of the non-gaussian nature of the
signal.  Because fluctuations in the linear density field are
gaussian, the power spectrum provides a complete characterization of
that field.  However, the ionized fraction is most assuredly
\emph{not} a gaussian random field; it is actually closer to a
discrete variable (taking values zero or unity) in the simplest
models.  During the middle to late stages of reionization, other
statistical properties may be better descriptors of the observations.
For example, \cite{furl-21cmobs} argue that such statistics can help
to distinguish the way in which reionization proceeds, whether from
high-to-low density regions or from low-to-high density regions.  More
detailed methods to extract this information from the measurements are
still required, however.

\section{Observational Challenges \& Instrument Requirements}

\subsection{Foreground Contamination}

The surface brightness fluctuations $\delta T \sim 1$--$10 \mkel$ in
the redshifted 21cm spectral line amount to ${\sim}10^{-5}$ of the
brightness of the radio sky, which is dominated by the diffuse
continuum synchrotron emission from the Milky Way Galaxy.  The
redshift range $6.2 < z < 20$ corresponds to $197 > \nu > 68 \Mhz$
[where $\nu=1420(1+z)^{-1} \Mhz$].  In this frequency range, a good
rule of thumb for the brightness temperature in the coldest directions
is $T_B\approx 180(\nu/180{\rm MHz})^{-2.6}$K.  The greatest
complication imposed by the Galactic foreground is the increase in
receiving system noise $T_{sys}$, since, according to the radiometer
equation, the $rms$ noise in the antenna temperature $\Delta T\sim
T_{sys}/\sqrt{\Delta\nu\;t_{int}}$ rises in proportion to sky
brightness ($T_{sys}\geq T_{sky}$); achieving a low noise level must
be accomplished by increasing integration time $t_{int}$, since the
bandwidth $\Delta\nu$ is set by matching the expected widths of the
spectral features (${\sim}0.2$ to 2 MHz).  The brightness temperature
of the radio sky increases toward the Galactic plane and is an order
of magnitude higher toward the Galactic Center, necessitating the
identification of selected high Galactic latitude regions in the sky
for studying 21 cm emission from the EOR.  The Galactic synchrotron
halo is considered to be smooth on arcminute scales when mapped at
normal sensitivities.  However, the halo's fluctuations at mK
sensitivity remain to be explored.

To first order, conventional radio spectroscopic techniques for
continuum subtraction should be capable of removing the foregrounds
created by the Galactic synchrotron and by discrete extragalactic
sources \cite{briggs04}. Interferometric antenna arrays are
insensitive to smooth emission, and ``resolve away'' the DC term of
the Fourier transform of the sky brightness.  This leaves behind the
structure caused by fluctuations in the Galactic foreground and the
discrete radio source populations on the sky.  Continuum source
spectra are typically either power laws or vary only slowly from a
power law. As a consequence, there are two steps to radio continuum
subtraction, depending on the strength of the contaminating sources.

In the first step, the coordinates and spectrum of the strongest
sources are measured, and the instrumental response to these sources
is subtracted directly from the calibrated $u$-$v$ dataset. This leaves
behind the many weaker sources, whose sum provides the continuum
emission that populates the 1 to 20 arcminute resolution pixels
in which we wish to measure the 21cm EOR emission.

\begin{figure}[htb]
\includegraphics[width=3in]{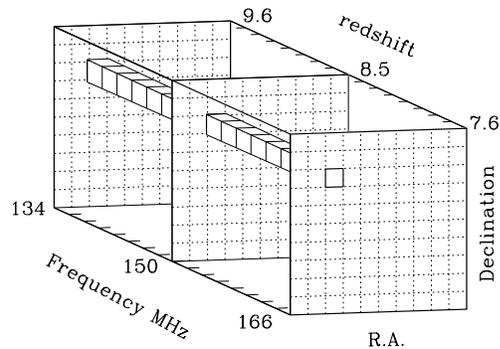}
\vspace{-0.5in}
\caption{Spectral line data cube. The spectral information for one
R.A, Declination coordinate has been drawn in for emphasis. Since
radio interferometers do not measure the `zero spacing flux density,'
the average flux density of each frequency plane of an observation
cube is equal to zero.  Spectral baseline removal involves fitting the
residual continuum for each R.A-Dec pixel separately with a low-order
polynomial. The fitted function is then subtracted from the values as
a function of frequency.  (from \cite{briggs04})}
\label{fig:cube}
\end{figure}

\begin{figure}
\includegraphics[width=3in]{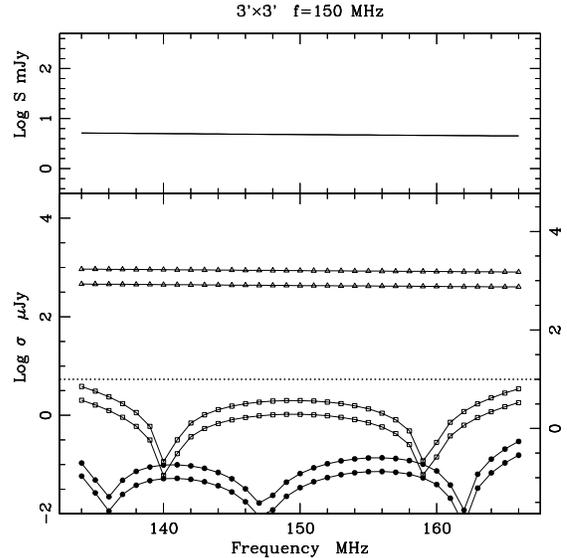}
\vspace{-0.5in}
\caption{ An illustration of spectral baseline foreground subtraction
for $3'{\times}3'$ pixels. Top panel: average integral flux density
vs. frequency per pixel; this includes radio point sources but assumes
that both the smooth Galactic contribution and the continuum sources
brighter than $S_{cut}=100$~${\mu}$Jy have been subtracted.  Bottom
panel: Open triangles are $rms$ deviations without continuum
subtraction. Open squares are $rms$ residuals after linear spectral
baseline subtraction. Filled circles are $rms$ residuals after
quadratic baseline subtraction.  In each case, the upper curve
includes fluctuations from large scale structure , while the lower
curve includes only Poisson statistical fluctuations (based on mean
source count statistics).  The horizontal dotted line marks the 10~mK
level at 150~MHz, which is typical of the $rms$ fluctuations expected
for the EoR signal. The righthand axis indicates $\log T_B$ for this
pixel size (from \cite{briggs04}).  }
\label{fig:subtr}
\end{figure}

The second step operates on a pixel by pixel basis.  The standard
continuum subtraction operates on the intermediate stage data-product
from spectral line observations, which is a data-cube (see
Figure~\ref{fig:cube}) composed of many image planes, with each plane of
different frequency.  The sums of the power law spectra with different
spectral indices that fill a R.A.-Declination pixel may have spectral
curvature, but these are adequately fit by second order polynomials.
Several authors \cite{DiMatt02,OhMack03,Cooray04,DiMatt04} have raised
concerns about the confusing effects of discrete source foregrounds,
but conventional techniques are adequate for their removal, as
illustrated in Figure~\ref{fig:subtr}.

\begin{figure}
\includegraphics[width=3in]{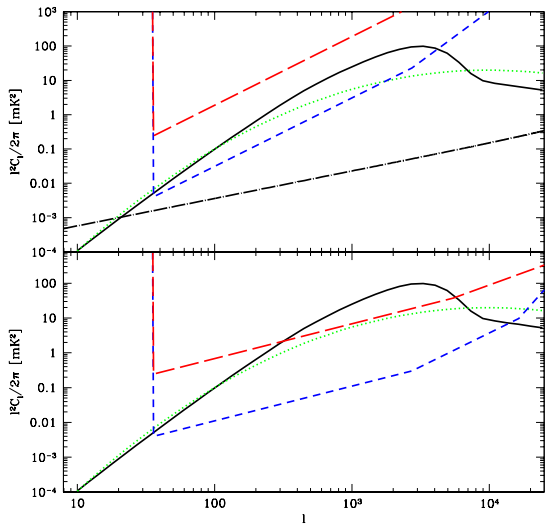}
\vspace{-0.5in}
\caption{Instrument sensitivities, as estimated by \cite{Zald04}.  In
  both panels the dotted and solid curves show simple estimates for
  the 21 cm signal with and without reionization, respectively.  The
  top panel shows the noise power spectrum (relevant for imaging) for
  LOFAR (long-dashed curve) and SKA (short-dashed curve).  The bottom
  panel shows estimates for the errors in logarithmic $l$-bins of a
  statistical measurement of the power spectrum.  The dot-dashed curve
  in the top panel shows the estimated level of residual foreground
  contamination from faint radio continuum sources. }
\label{fig:noise}
\end{figure}

Other approaches to the problem of continuum removal have also been
studied.  They too rely on the spectral smoothness of the
contaminants.  One way to approach it is through the symmetries in the
data cube in Fourier space \cite{Mor04}.  Another is through frequency
differencing \cite{Zald04}, where neighboring planes in the data cube
are subtracted from each other.  This technique is similar to that
described above except that the baseline subtraction is performed
implicitly on the visibilities.  The top panel of
Figure~\ref{fig:noise} illustrates the level to which the foregrounds
can be removed in the formalism of \cite{Zald04}.  The solid and
dotted curves show estimates of the 21 cm fluctuation spectrum at
$z=10$, with and without HII regions.  The dot-dashed curve shows the
level to which the radio point source population predicted by
\cite{DiMatt02} can be removed.  It is clear that smooth foregrounds
allow for more than the necessary precision in continuum removal.

Additional uncertainties surround the subtleties of instrumental
response to the known (and yet to be discovered) foregrounds. For
example, an array telescope actually senses \emph{every} continuum
source above the horizon at some level; the response has strong
frequency dependence that increases with angular separation of each
source from the telescope pointing coordinates. The distribution of
sources (in strength and location) produces a pseudo-random
superposition of these effects in the net response and could create
spurious spectral features that will confuse 21cm line studies of the
EOR.  The solution may be to perform the data reduction for EOR
surveys with access to all-sky radio source catalogues including
coordinates, flux densities and spectral shapes. Another subtlety
arises from the unknown polarization properties of the Galactic halo
foreground, in which differential Faraday rotation of the synchrotron
emission will enter in a frequency dependent way. If the telescope
polarization response is not calibrated to high precision, then the
rotating polarization vector could couple in the analysis to the
measured intensity, and this could vary with frequency and position in
the sky to mimic EOR signals.  Much remains to be done in the
exploration of instrumental effects, astrophysical foregrounds, and
their interactions within the observational process.

\subsection{SKA requirements}

The goal of observing 21 cm emission from the EOR constrains the
frequency coverage and array configuration of the telescope. For
example, the SKA \emph{must} be able to observe at frequencies on the
order of $100 \Mhz$.  The opacity in the Gunn-Peterson trough in high
redshift quasars suggests that reionization may be ending at $z \sim
6$ \cite{fan}.  If the universe is largely neutral ($\bxh > 0.3$) up
to this epoch, even limiting the frequency coverage to $>150 \Mhz$
will provide valuable information about reionization.  On the other
hand, the CMB data presently indicates that reionization must begin at
$z \sim 15$ \cite{kogut03}.  If the universe is partially ionized
beyond this point, the SKA would need to push to much lower
frequencies in order to make 21 cm measurements.  Unfortunately,
precise constraints on the reionization history will probably not
appear for several years (see \S 5) and may even be best determined by
21cm line observations.  In any case, it appears that measurements of
the formation of the first structures, as described in \S3.1--3.2,
could require frequency coverage below $70 \Mhz$.

\begin{figure*}
\includegraphics*[width=6in]{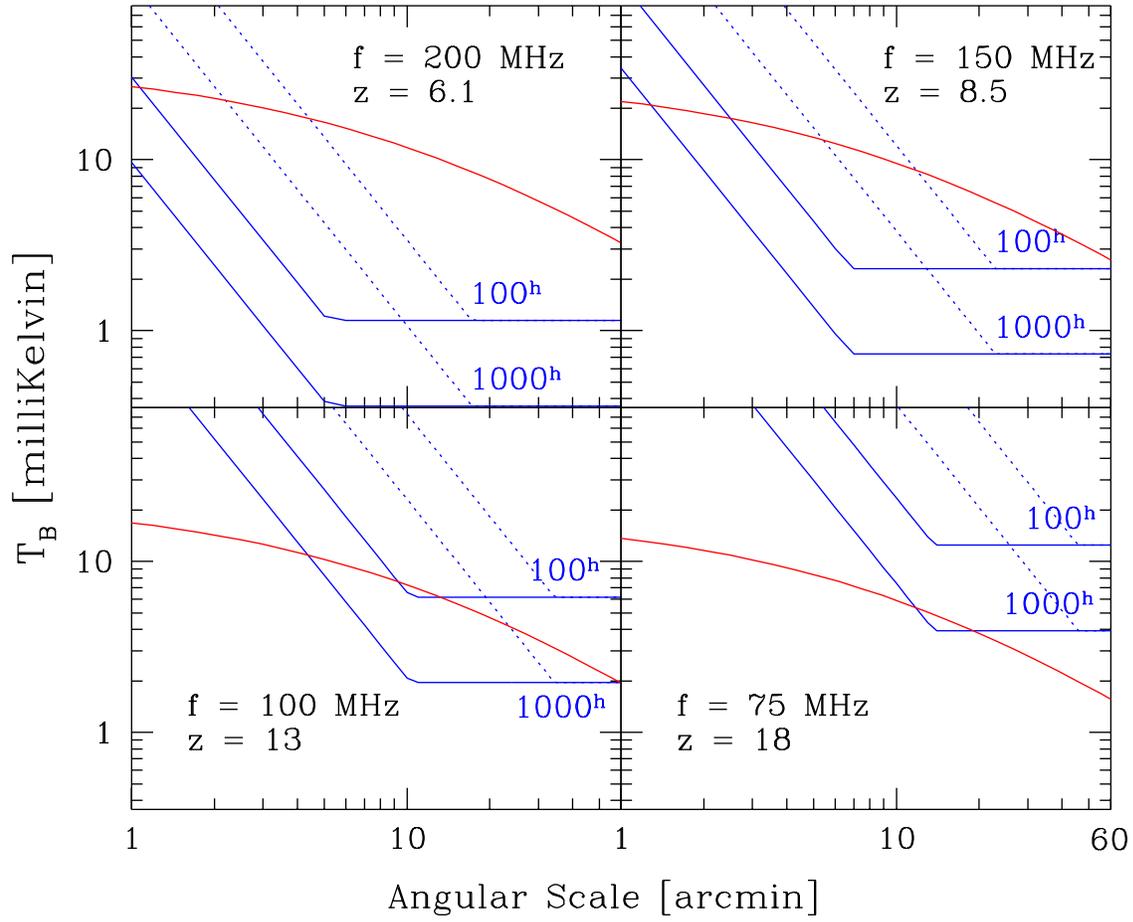}
\vspace{-0.3in}
\caption{Estimated detection limits of faint EOR features at four
redshifts. The brightness fluctuation spectrum for a ${\Lambda}$CDM
model assuming pure density fluctuations (following
\cite{furl-21cmsim}) has been scaled upward by a factor of three to
provide a rough measure of the 3-$\sigma$ peaks of brightness. We
assume $\Delta \nu=0.5 \Mhz$ and neglect redshift space distortions.
Solid lines indicate 5$\sigma$ detection limits for 100 and 1000 hour
SKA observations, for different diameter arrays obtained by diluting
1~km$^2$ of collecting area to achieve higher angular
resolution. Dotted curves do the same for a LOFAR array.}
\label{fig:ska_sens}
\end{figure*}

Sensitivity to faint fluctuations in surface brightness against the
bright radio sky at these frequencies will require a large collecting
area.  The achievable uncertainty in surface brightness
${\Delta}T_b\approx \Delta T_{a}/\eta_f =
T_{sys}/\eta_f\sqrt{\Delta\nu\;t_{int}}$ for emission that fills the
array beam depends on the array filling factor $\eta_f$. Because
the fluctuation spectrum falls off toward large angular scales, the
array must have diameter of $\sim$3000 wavelengths if the beam is to
match $1'$ structure. On the other hand, to obtain surface brightness
sensitivity, the aperture cannot be too severely diluted due to the
$\eta_f$ dependence.

The filling factor requirement, along with the frequency dependence of
the radio sky, are illustrated in Figure~\ref{fig:ska_sens}. This
Figure provides an overview of the issues that influence the ability
of the SKA to map specific features in the EOR, which will be crucial
for relating structures in the IGM to star forming regions observed at
infrared and possibly millimeter wavelengths. The Figure has four
panels in order to show the increasing degree of difficulty of the
observation toward higher redshifts: as the frequency decreases, the
sky brightness and the diffraction limited resolution worsen.
The signal predictions assume pure density fluctuations (i.e., they
neglect the contribution from HII regions) and that $T_S >> T_{CMB}$
everywhere.  They are generated as described in \cite{furl-21cmsim}
(who in turn followed \cite{Tozzi00}), assuming $\Delta \nu=0.5 \Mhz$
and neglecting redshift space distortions (which will amplify the
signal by a small factor).  Note that we show $3\sigma$ peaks in order
to illustrate the brightest features.
The instrumental sensitivity is best for the largest
angular scales ($>10'$) since the array beam would be filled by the
brightness fluctuations for a filled array ($\eta_f=1$). For finer
angular scales, the aperture must be diluted in order to match the
beam to the fluctuations, causing the sensitivity to deteriorate.
For this figure, the noise level is set by the $T_{sys}\approx T_{sky}$,
with the assumption that residuals from foreground subtraction make a
negligible contribution.

The statistical measures of the fluctuation power spectrum have
somewhat different dependencies.  The short-dashed curve in the upper
panel of Figure \ref{fig:noise} shows the approximate noise power
spectrum for the SKA assuming one month of continuous observing, a
$0.4 \Mhz$ channel, and $A_{\rm eff}/T_{\rm sys}=2 \times 10^5 {\rm
m}^2/\kel$ (see reference \cite{Zald04} for details).  This curve
essentially shows the noise \emph{per visibility} as a function of
scale and is analogous to the noise per pixel in a real-space map.
The long-dashed curve shows the approximate noise for the \emph{Low
Frequency Array} (LOFAR).  We can see clearly that (because the sky
brightness is essentially fixed) high signal-to-noise maps require
collecting areas on the order of a square kilometer; even then they
will be difficult.  We note that this sensitivity curve was
constructed with several fairly crude approximations and is only
accurate to a factor of a few.  It depends in detail on the
configuration of the array, the antenna beams, the correlator
properties, and the observing strategy.  In the future, these
questions must be addressed in more detail.  Also note that, on the
scales of the HII regions, the contrast between ionized and neutral
regions will be substantially larger than the simple \emph{rms} variation
shown here; in this sense map-making during the middle stages of
reionization will be easier than suggested by this plot.

Of course, statistical properties are much easier to measure than
maps, because we can average over many measurements on each scale.
The requirements for the signal-to-noise ratio on each angular scale
are therefore much weaker.  The bottom panel of Figure~\ref{fig:noise}
shows the estimated errors in such a measurement for the SKA and
LOFAR, assuming logarithmic bins in $l$-space and a $100$ square
degree field of view.  We see that the power spectrum of the
fluctuations can be measured with good accuracy even with collecting
areas smaller than a square kilometer \emph{provided} that the field
of view is large.

Finally, the most important angular scales for measuring the
fluctuations are in the range $\sim 1$--$20'$, which contain both the
peak of fluctuations in the linear density field and the expected
scale of HII regions during most of reionization (at least in the
model of \cite{furl-21cmps}).  This suggests that the SKA have a
compact core with baselines $\sim 3 \km$ that contains a substantial
fraction of the collecting area (Figure~\ref{fig:noise} assumes that
$20\%$ of the antennae are inside a core of diameter $3 \km$).  While
long baselines will be useful for point source removal, 21 cm
tomography requires heavy emphasis on the compact core.

\section{Conclusions}

Understanding reionization is one of the major goals of
cosmology.  We have argued that 21 cm tomography is a particularly
powerful way to attack the problem, but there are several others.  We
are already reaching the limits of observations with the Ly$\alpha$
forest; observations of complete Gunn-Peterson troughs at $z \sim 6$
indicate that this tool cannot be taken to substantially higher
redshifts.  Moreover, Ly$\alpha$ absorption is so strong that even
measurements of zero transmission do not imply a fully neutral medium.
Thus, more detailed constraints from optical observations must rely on
modeling of the HII region around quasars \cite{Wyithe04a,Mes04},
measurements of the detailed shape of the Ly$\alpha$ absorption
profile \cite{MirEsc98} in (for example) gamma-ray burst afterglow
spectra, or the careful interpretation of Ly$\alpha$ emission lines
from high-redshift galaxies
\cite{pello04,loeb04,ricotti04,gnedin04,cen04}.  While invaluable,
these methods are fraught with uncertainty.

More detailed measurements of the CMB will also yield more information
about the ionization history.  Most importantly, \emph{WMAP} and
\emph{Planck} will better measure the total optical depth to electron
scattering and place a strong integral constraint on the ionization
history.  However, the CMB power spectrum is relatively insensitive to
the evolution of the ionized fraction \cite{holder03}, and strong
constraints will probably have to await space-borne missions beyond
\emph{Planck}.  (Note that the reionization signal appears on larger
angular scales than ground-based or balloon experiments can easily
probe.)

Finally, direct observations of high-redshift protogalaxies, with (for
example) the \emph{James Webb Space Telescope} or the upcoming
generation of twenty-meter ground-based telescopes, will obviously
teach us an enormous amount about the growth of the first luminous
objects.  However, we will still have only indirect measurements of
the IGM structure and the process of reionization.

Thus it seems that 21 cm tomography provides a unique and invaluable
probe of the high-redshift universe.  No other technique can measure
the three-dimensional distribution and time evolution of neutral gas
in such detail.  Also, only 21 cm measurements can extend into the era
\emph{before} the formation of the first luminous objects (see \S 3.1
and 3.2) to map the initial stages of structure formation.  Provided
that the technical challenges described in \S 4 can be overcome,
there is no doubt that 21 cm tomography will revolutionize our
understanding of reionization and the early universe.  While
statistical measurements of the distribution of ionized gas may be
possible with LOFAR or some other instrument before SKA, we expect
that a collecting area near one square kilometer will be required in
order to make high signal-to-noise maps of the neutral gas
distribution.  

On the other hand, we expect that these measurements will complement
the next generation of infrared, X-ray, and CMB instruments.  For
example, comparing the distribution of ionized gas to the luminous
sources in a cosmological volume would show explicitly how those
sources are responsible for ionization.  We can also use 21 cm
observations of large ionized regions to locate bright quasars and in
principle to constrain many of their unknown properties, including
their ages \cite{Wyithe04b} and spectra \cite{Tozzi00}.

Finally, we note that reionization is an evolving field.  The data are
continuing to improve, and our theoretical understanding is advancing
even more rapidly.  At the same time, the amount of attention focused
on 21 cm studies has expanded dramatically, and our understanding of
the instrumental and data analysis challenges is improving by leaps
and bounds.  This review has presented a snapshot of the field in May
2005.  Our expectations will no doubt change in many ways (some
unexpected) before the SKA begins observations.

\end{document}